\documentclass[12pt,preprint]{aastex}
\newcommand{\gtsim}{\raisebox{-1.0ex}{$\stackrel{\textstyle>}{\sim}$}}
\newcommand{\ltsim}{\raisebox{-1.0ex}{$\stackrel{\textstyle<}{\sim}$}}
\def\kms{km~s$^{-1}$}

\def\goes{{\sl GOES}}
\def\yohkoh{{\sl Yohkoh}}

\def\hinode{{\sl Hinode}}

\def\p78{{\sl P78-1}}

\def\stereo{{\sl STEREO}}
\def\sdo{{\sl SDO}}

\def\fexii{Fe~{\sc xii}}
\def\fexv{Fe~{\sc xv}}

\def\heii{He~{\sc ii}}

\def\kms{km~s$^{-1}$}

\def\etal{et~al.}

\begin{document}
%


\slugcomment{Accepted 2018 Nov 7: \underline{The Astrophysical Journal}}

\title{A Two-Sided-Loop X-Ray Solar Coronal Jet Driven by a Minifilament Eruption}

\author{Alphonse C.~Sterling\altaffilmark{1}, Louise K. Harra\altaffilmark{2}, Ronald L. Moore\altaffilmark{1,3}, 
\& David A. Falconer\altaffilmark{1,3}}

\altaffiltext{1}{Marshall Space Flight Center, Huntsville, AL 35812, USA;
alphonse.sterling@nasa.gov}
 
\altaffiltext{2}{UCL-Mullard Space Science Laboratory Holmbury St Mary, Dorking, Surrey, RH5 6NT, UK; l.harra@ucl.ac.uk}
 
\altaffiltext{3}{Center for Space Plasma and Aeronomic Research, University of Alabama in Huntsville, 
Huntsville, AL 35899, USA; ron.moore@nasa.gov,David.a.Falconer@nasa.gov}

\begin{abstract}

Most of the commonly discussed solar coronal jets are of the type consisting of a single spire extending approximately 
vertically from near the solar surface into the corona.  Recent research supports that eruption of a miniature filament
(minifilament) drives many such single-spire jets, and concurrently generates a miniflare at the eruption site.  A
different type of coronal  jet, identified in X-ray images during the \yohkoh\ era,  are {\it two-sided-loop jets}, which
extend from a central excitation location in opposite directions, along low-lying coronal loops more-or-less horizontal
to the surface.  We observe such a two-sided-loop jet from the edge of active region (AR)~12473, using data from \hinode\ XRT
and EIS, and \sdo\ AIA and HMI\@.  Similar to single-spire jets, this two-sided-loop jet results from eruption of  a
minifilament, which accelerates to over 140~\kms\ before abruptly stopping after striking overlying nearly-horizontal
loop field at $\sim$30{,}000~km altitude and producing the two-sided-loop jet.  Analysis of EIS raster scans show that 
a hot brightening, consistent with a small flare, develops in the aftermath of the eruption, and that Doppler motions 
($\sim$40~\kms) occur near the jet-formation region.  As with many single-spire jets, the magnetic trigger 
here is apparently flux cancelation, which 
occurs at a rate of $\sim$4$\times 10^{18}$~Mx/hr, comparable to the rate observed in some single-spire AR jets.  An apparent
increase in the (line-of-sight) flux occurs within minutes of onset of the minifilament eruption, consistent with the
apparent increase being due to a rapid reconfiguration of low-lying field during and soon after minifilament-eruption onset.

\end{abstract}


\section{Introduction}
\label{sec-introduction}

Coronal jets are transient collimated ejections of solar material, typically of 
length $\ltsim$10$^5$~km and width $\sim$10$^4$~km.
They were first observed in detail in X-rays with the \yohkoh\ Soft X-ray Telescope (SXT)
\citep{shibata_et92,shimojo_et96}.  Later they were studied with the X-Ray Telescope (XRT) on \hinode\
\citep{cirtain_et07,savcheva_et07}.  More  recently they have been studied in EUV with the \stereo\
spacecraft \citep[e.g.][]{nistico_et09}, and with the Atmospheric Imaging Assembly (AIA) on
Solar Dynamics Observatory (\sdo) \citep[e.g.][]{moore_et13,sterling_et15}.  There are many
additional jet studies with a variety of instruments \citep{raouafi_et16}.

Typical jets consist of a bright base near the solar surface,  with a single spire extending
progressively  outward from near the solar surface into the corona. 
(\citeauthor{shibata_et94}~\citeyear{shibata_et94} called these ``anemone jets''; here we will 
refer to them as ``single-spire jets.'') It was first suggested that such jets result according  to
an ``emerging-flux model,'' whereby a magnetic bipole emerges from below the solar surface into the
corona and undergoes  magnetic reconnection with surrounding open (or far-reaching) nearly vertical
field, with the  jet spire forming along that open field
\citep[e.g.,][]{shibata_et92,yokoyama_et95}.  Later however, improved resolution and  wavelength
coverage with AIA showed that many (if not most) jets instead result from eruption of a  miniature
filament, or  ``minifilament,'' of size scale $\sim$10{,}000~km
(\citeauthor{sterling_et15}~\citeyear{sterling_et15}; also see, e.g.,
\citeauthor{shen_et12a}~\citeyear{shen_et12a}, \citeauthor{adams_et14}~\citeyear{adams_et14},
\citeauthor{panesar_et16}~\citeyear{panesar_et16}), accompanied by a ``miniflare'' brightening at
the edge of the jet base (this brightening is  sometimes called a jet(-base) bright  point, or JBP;
e.g.\ \citeauthor{sterling_et15}~\citeyear{sterling_et15}).  Meanwhile,  improved magnetic field
coverage with \sdo's Helioseismic and Magnetic Imager (HMI) showed that often jets coincide with
magnetic flux cancelation \citep[e.g.,][]{huang_et12,young_et14,adams_et14}; in some cases where
flux emergence coincides with the jets, the jets still originate from locations where one pole of
the emerging bipole is canceling with surrounding field \citep[e.g.][]{shen_et12a,li_et15}.  From
the ``minifilament-eruption-model" standpoint, the flux cancelation builds the minifilament-holding
magnetic field \citep{panesar_et17}, and triggers it to erupt \citep{panesar_et16,panesar_et18} to
form the jet.

In addition to single-spire jets, there are also {\it two-sided-loop} coronal jets 
\citep{shibata_et94}, whereby two spires develop roughly symmetrically and horizontally to the
surface from a central bright region.  These two-sided-loop jets were first seen in coronal X-ray
images, and are also seen in EUV images \citep[e.g.][]{alexander_et99,jiang_et13}.  The
emerging-flux model was also invoked to explain these jets theoretically, with the emerging flux
reconnecting with overlying horizontally  directed field, and numerical simulations of this showed
results similar to the \yohkoh/SXT X-ray observations \citep{yokoyama_et95,yokoyama_et96}. 
Published examples of \sdo-era observations, with high-resolution and high-cadence multi-wavelength
images and magnetograms, of  these two-sided-loop jets are sparse  (see however
\S\ref{sec-discussion} for recent references). Here we present \hinode\ and \sdo\ observations 
showing strong evidence that a two-sided-loop jet resulted from an erupting minifilament, similar
to that in single-spire jets.

\section{Instrumentation and Data}
\label{sec-data}

Our observed two-sided-loop jet occurred on 2015 December~30 near 22:41~UT, southeast of a set 
of sunspots in NOAA active  region (AR) 12473 that was located at
heliocentric  latitude and longitude of about -20, +45, producing a C-level
enhancement in the  \goes\ X-ray flux.

We use imaging data from both XRT and AIA\@.  We confirm that it is an  ``X-ray jet'' with  XRT,
assuring that we are studying a two-sided-loop jet similar to those identified by 
\citet{shibata_et94} in X-rays with SXT\@.  XRT has spatial pixel resolution of $1''.02$, with
variable time cadence and field of view (FOV) \citep{golub_et07}.  AIA produces full-Sun images
in seven EUV bands at 12~s cadence with $0''.6$~pixels \citep{lemen_et12}; for this study we examined all
EUV channels, and the 1600~\AA\ UV channel.  For our purposes here, we find it
adequate to present 304~\AA, 193~\AA, 211~\AA, and 94~\AA\ images (and their 12-s-cadence movies), which 
respectively have peak contributions at $5\times 10^4$, $1.6\times 10^6$, $2.0 \times 10^6$~K, and 
$6.3 \times 10^6$~K\@.  We also use
raster-image data from the \hinode/EUV Imaging Spectrometer \citep[EIS,][]{culhane_et07} in the
\heii~304~\AA, \fexii~195~\AA, and \fexv~284~\AA\ lines.  Each scan required $\sim$3.5~min to
cover an E-W extent of $\sim$50$''$, with a N-S slit extent of $\sim$120$''$.  There  was a
fortuitous overlap with the jet event in the southwest quadrant of EIS's limited FOV\@.  We also use
magnetograms from HMI \citep{scherrer_et12}, which has maximum cadence 45~s and $0''.5$ pixels.

\section{Observational Results}
\label{sec-results}

\subsection{X-Ray and EUV Evolution}
\label{subsec-images}

Figures~1(a---b) show the two-sided-loop jet in XRT images with classic morphology 
\citep{shibata_et94,yokoyama_et95,yokoyama_et96} in X-rays, with a strong brightening in-between
the two loop-confined arms (spires) of the jet; the 
accompanying video shows the jet's evolution. These \hinode/XRT images are very close in appearance to the
snapshot from \yohkoh/SXT in \citet{yokoyama_et95} (see Fig.~3(b) of that paper; the figure is also in 
\citeauthor{shibata_et94}~\citeyear{shibata_et94}).  Both the \citet{yokoyama_et95} 
jet and our jet have one side that is smaller and brighter (to the south in the 
\citeauthor{yokoyama_et95}~\citeyear{yokoyama_et95} jet, 
and to the northeast in ours), and one side that is larger and dimmer (north in 
\citeauthor{yokoyama_et95}~\citeyear{yokoyama_et95}, and southeast in ours).
Viewing the movie accompanying our Figure~1(a---b), the frame at 22:46:10~UT comes closest in appearance to
the snapshot in the \citet{yokoyama_et95} figure (we choose to show different frames in our Fig.~1(a,b) 
because of the strong saturation of the flaring location in that 22:46:10~UT frame).  As with the original observations with \yohkoh\ 
however, the X-ray images alone yield little direct information on the cause of the jet.  We look
to additional data for more insight. 


Figures~1(d---f) show the jet 
with AIA~94~\AA\ images, with an initial brightening in Figure~1(d) (also visible in X-rays in Fig.~1(a)) 
the two-sided-loop 
jet structure in Figure~1(e), and a closeup in~1(f).  Arrows in 1(e) point to oppositely directed flows
along two strands of the two jet-guiding lobes.  Figures~1(g---h) show the
event in AIA~193~\AA, with arrows in 1(g) showing the elevated field that later becomes one of the
jet-guiding lobes.  Figure~1(i) shows  
a closeup of the region that brightens at the earliest sign of jet activity, with
a filament-like absorbing feature (arrows) that the accompanying video shows to be erupting. 
The erupting segment (black arrow) has length of  $\gtsim$15$''$ ($\approx$11{,}000~km).  This
is similar to the size (8000~km) \citet{sterling_et15} found for erupting minifilaments when
they were at about the same relative distance above the surface as the in Figure~1(i). This is smaller 
than the sizes of typical filaments \citep[$\sim$30{,}000---110{,}000~km;][]{parenti14}. Thus, 
just as with the more typical ``single-spire'' jets, this two-sided-loop jet apparently results from 
an erupting minifilament. From the accompanying videos, the jet forms when this erupting 
minifilament collides with pre-existing overlying field shown by arrows in Figure~1(g).

Figure~2 shows AIA~94~\AA\ images with a larger FOV than Figure~1, thereby revealing 
more clearly the overall jet structure.  
After an initial brightening (red arrow in Figs.~2(a,c)), the jetting into the two 
lobes eventually extends out 
asymmetrically from that brightening, longer toward the southeast and shorter toward the 
northwest (Fig.~2(c)).  We see the southeast
lobe develop in time in a similar fashion in several AIA wavelengths; in these 94~\AA\ images the
jet-spire brightening in the lobe indicated by the white arrow in Figure~2(b) extends away from the initiation
site at $\sim$550$\pm$100~\kms.  In contrast, the northeast lobe appears to brighten mostly in place,
and therefore seems to be mostly a result of material entering into the passband
of sensitivity for the 94~\AA\ wavelength channel as it heats or cools, rather than 
arising from a front propagating along the field to fill the loop.
Because of the differences in temperature response in the 94~\AA\ EUV of AIA \citep{lemen_et12} 
and the broad-band X-rays of XRT \citep{golub_et07}, jets can appear differently in the
different wavelength regimes \citep[cf.][]{sterling_et15}.  This is the case here, with for example
the northwestern lobe of the jet clearly visible from the earliest XRT images (22:44:09~UT), while
it does not appear until substantially later in 94~\AA\ (orange arrow in Fig.~2(c)).  Evolution 
of the southwest lobe is similar
in X-rays to that in 94~\AA, with flows along the lobe field from the central region visible over
22:44---22:48~UT, approximately coinciding with the flows visible in the 94~\AA\ images.  

Two-sided-loop jets (and single-spire jets too, for that matter) were
originally defined from X-ray images, and thus the images expected to match most closely those
in the sketches of \citet{shibata_et94} and \citet{yokoyama_et95,yokoyama_et96} are those of Figure~1(a)
and~1(b), although unfortunately this appearance is compromised somewhat by the limited
FOV of our XRT images in Figure~1.

Figure~3 shows a zoomed-in view of the erupting minifilament, this time in AIA 304~\AA\ images.
Over some time the minifilament becomes visible near the surface (Fig.~3(a)). Figure~3(b) shows 
a brightening beneath the minifilament as it has just started to rise; this brightening is visible in
all AIA channels, and so includes hot emissions.  A short time later (Fig.~3(c)), the brightenings
have become much more pronounced.  Moreover, the minifilament itself now shows a contorted, serpentine-like
shape -- likely due to writhing -- as it erupts away from the surface (Fig.~3(d)). At the time
of Figure~3(e), the erupting and expanding minifilament reaches the pre-existing overlying field shown
by the white arrows in Figure~1(g), and by light blue arrows here in Figures~3(d---f).
From the corresponding movie, upon pushing into the overlying field, the minifilament appears to show 
untwisting motions over approximately 22:44---22:48~UT.  Along with and continuing after this unwinding, 
flows toward the northwest occur from the erupting minifilament (green arrow in Figs.~3(e---f)).  Also,
the overlying feature (light-blue arrows) shows distinct flows southeastward from approximately the
time it is impacted by the erupting minifilament field, at about 22:45~UT, and continuing until about 22:51~UT
in the movie.  These features are also visible in the other AIA EUV channels; see the videos accompanying
Figure~1.



Figure~4 shows the trajectory of the erupting minifilament as a function of time, where we have visually
estimated the top of the minifilament as it moves upward in 211~\AA\ images (Fig.~1(i)).   We perform each
length measurement three independent times to estimate the random uncertainty in the measured value.  
Upward movement of the central part of the filament commences near 22:34~UT, indicated by the 
upward-pointing arrow. This is the 
time of the initial brightening at the base of the minifilament indicated by the green arrow in
Figure~3(b); it is also visible in other AIA videos accompanying Figure~1, for example in 94~\AA\ video 
of Figure~1(f) from 22:34:24, and especially at 22:35:24~UT\@.  In Figure~4 the green curve represents 
the AIA 94~\AA\ flux integrated over the region of the box in Figure~1(d), but the 22:34:24 UT flux 
increase visible in the video is too weak, relative to the background  intensity, to stand out in the 
green light curve.  (The
enhanced background intensity in 94~\AA\ images, visible as a hump in the green curve of Fig.~4 that
peaks  near 22:30~UT, is caused by other dynamical processes that we do not investigate here.)  From
about 22:40~UT there is a sharp increase in the rise trajectory of the minifilament, and this corresponds
to an increase in the 94~\AA\  integrated intensity (down-pointing arrow in Fig.~4, visible over 22:39---22:41~UT
in the video accompanying Fig.~1(f)), and then a further
acceleration of the minifilament near 22:42~UT, indicated by the velocity plot (orange) in Figure~4,
which is also accompanied by a sharp increase in the integrated 94~\AA\ intensity.  This latter-most
sharp  increase in the 94~\AA\ intensity can be considered to be the growth of a small flare 
arcade accompanying the accelerated rise of the minifilament.

This behavior of the minifilament's rise, with slow-rise phases prior to the 
most-accelerated rise, EUV brightening enhancements accompanying accelerations in the rise
trajectory, and onset of strongest flaring accompanying the strongest minifilament upward
acceleration, all closely mimic behavior of larger-scale erupting filaments and their
accompanying flares 
\citep[e.g.,][]{sterling_et05,sterling_et07,sterling_et14,imada_et14,mccauley_et15,harra_et17,green_et18}.


From Figure~4,
the minifilament erupts outward until it reaches a height (displacement) of about 23{,}000~km, when it  abruptly
stops, apparently arrested by overlying field that is visible even prior to the jet (arrows in
Fig.~1(g)).  Due to the projection angle, the true height might be $\sim$30\% larger than these 
plane-of-sky-projected values, as the event occurred at an angle of $\sim$45$^\circ$ with the Earth-Sun
line-of-sight; so the height is $\sim$30{,}000~km.  Comparison with previous studies of  ejective
eruptions \citep[e.g.,][]{sterling_et14} suggest that the minifilament's rise is relatively
unimpeded until the abrupt velocity decrease near 22:44~UT, when videos accompanying Figure~1 indicate
that it pushes into overlying magnetic field.

\subsection{EIS Rasters}
\label{subsec-eis}

Figure~5 shows an EIS raster intensity image in \heii\ (Fig.~5(a)), and 
a ratio of \fexv-to-\fexii\ intensity images (Fig.~5(b)).  These ratio
images provide a qualitative measure of hotter (brighter) and cooler (darker) locations in the region 
\citep{doschek_et07}.  Figure~5(c) shows contours from Figure~5(b) on an AIA 193 image.  The rectangle 
in Figure~1(i) shows that the EIS FOV covers only a small portion of the jet's base region. 

Figure~5(a) shows a dark feature (light-blue arrow) and a bright feature (black arrow), which  
correspond to cooler and hotter locations in Figure~5(b).  Comparing these features with the 
AIA features in the zoomed-in videos corresponding to Figures~1(f), 1(i), and 3 
suggests that the dark feature could be a void; see for example the Figure~3 304~\AA\ video 
between 22:42 and 22:45~UT, where there seems to be a region comparatively free of emitting material 
between the erupting minifilament and the photosphere at the location of the EIS dark feature.
For the bright feature (black arrow in Fig.~5(a)), comparison with the same videos shows convincingly
that this region corresponds to flare loops that develop below the erupting minifilament; see for example
the 211~\AA\ movie corresponding to Figure~1(f) from about 22:50~UT, when the features at this location
take on the distinct appearance of post-flare loops.  Unfortunately however, the
EIS scans did not capture the erupting minifilament at a time when it was most clearly visible in the
AIA images, such as in Figure~1(i).  There is only one EIS scan coinciding with times where AIA
shows the minifilament very clearly in the EIS FOV; that at 22:42:18~UT\@.  But that scan has a 
data dropout near the jet region, and so we are not confident in that scan.  At the time of 
the prior raster (22:38:44~UT), the minifilament has not yet started lifting off, and by the 
time of the next scan (22:45:52~UT), hotter material envelopes
the minifilament, apparent in the AIA videos accompanying Figure~1.  So what we can say with certainty 
from the EIS rasters is that they show: (a) brightening beneath the erupting minifilament, 
(b) temperature-ratio maps showing hotter plasma at the expected small-flare location, and
(c) Doppler velocities from single-Gaussian fits in several rasters show redshifts, and one raster 
(beginning at 22:45:52) shows a strongly blueshifted feature too (both red- and blueshifts have
maximums of $\sim$40~\kms); we consider these line shifts further
in~\S\ref{sec-discussion}.

\subsection{Magnetic Evolution}
\label{subsec-b}

Figure~1(c) shows an HMI magnetogram of the region, and Figure~1(h) shows it overlaid onto
an AIA 193~\AA\ image.  From the latter panel and accompanying video, the minifilament eruption
clearly occurs on a magnetic neutral line.  

Figure~6 shows the positive-polarity flux summed over the box in Figure~1(c) for the 36 hrs beginning at
0~UT on December~30.  There is a general flux decrease over that time period of  $\sim$4$\times
10^{18}$~Mx/hr, in agreement for flux-cancelation rates found in  (single-spire) AR jets \citep[see
Table~1 of][]{sterling_et18}.  There is however a prominent jump of about $4\times 10^{19}$~Mx over
$\sim$15~min, beginning nearly exactly with the onset of the minifilament's eruption.  This does not have
the appearance of flux emergence; if it were emergence, the emergence rate during the jump
would be at or above $\sim$1$\times 10^{20}$~Mx/hr, which is the approximate maximum  rate of AR growth
\citep{zwaan87}; the magnetograms show no indication of such large emergence.  Instead we suspect that we
are seeing an increase in the line-of-sight component of the field as the field vector at the surface 
changes its direction as the minifilament erupts; the positive-polarity roots of that field were oriented 
such that in response to the
eruption, the direction of the field turned toward our observation direction, perhaps a natural
consequence of the our perspective of viewing the region near the west limb (see discussion of
Fig.~7 below). Similar magnetic rapid
reconfigurations have been seen in flares \citep[e.g.,][]{moore_et84,wang_et94}. Also supportive of
this hypothesis is the continuation of flux decrease with time for more than 12 hrs following the jet
(Fig.~6). Hence, the evidence is that flux cancelation triggered the minifilament's eruption that drove
the two-sided-loop jet, consistent with flux cancelation triggering many or most single-sided jets
\citep[e.g.,][]{panesar_et16,panesar_et17,panesar_et18,sterling_et16,sterling_et17}.


\section{Summary and Discussion}
\label{sec-discussion}

We find a two-sided-loop jet to result from eruption of a miniature filament.  This is consistent with 
earlier observations that minifilament eruptions cause single-spire jets.  In this case, the erupting
minifilament reached a velocity of $\sim$140~\kms\ prior to impacting and reconnecting  with an
overlying, roughly horizontal portion of a sigmoid coronal-loop field at a  height of
$\sim$30{,}000~km.  Reconnection between the erupting minifilament (more specifically: the 
erupting-minifilament flux-rope field that holds the cool minifilament material) and that 
horizontal-loop field resulted in heating at the reconnection location, and expulsion of X-ray/EUV jets
in  both directions along the loop, producing the two-sided-loop jet.  EIS raster scans confirm
accentuated  heating at the suspected small-flare location, and line-of-sight magnetogram observations 
are consistent with flux cancelation being the  trigger for onset of the minifilament's eruption.

Figure~7 shows a schematic of our minifilament-eruption scenario for producing this two-sided-loop
jet.  The minifilament flux rope writhes as it erupts, allowing reconnection at the elevated location
(Fig.~7(b)).   This schematic is topologically the same as that drawn for single-spire jets in
\citet{sterling_et15}, but with the ``open'' field now the long horizontal loop, running from 
right-to-left in the figure (northwest-to-southeast in the observed jet in Fig.~1).

Figure~7 also illustrates, with the purple arrows, our idea for the positive-flux  jump at and
soon after the time of minifilament-eruption onset.  These arrows represent the direction  of the
positive component of the photospheric field inside the footpoint of the minifilament loop (in the
figures, for  clarity we draw the arrows just outside of the minifilament loop).  Prior to the eruption
of the minifilament, the direction of the field vector in the minifilament's positive-polarity flux
patch is approximately vertical to the surface. Within a few minutes following the 
minifilament-eruption's onset however, the
foot of the positive-polarity field that undergoes the flare-arcade-building reconnection points more
toward the upper  left of the schematic: In other words, after the eruption's onset, the purple arrow 
in Figure~7(b) is no longer vertical, as it was in Figure~7(a); rather, it is now leaning slightly couter-clockwise
to the normal to the surface, pointing to the southeast (in the plane of the figure).  This is a 
result of the reconnecting field beneath
the erupting minifilament collapsing in upon itself. (We speculate that this photospheric-field-direction
change could be a byproduct of the so-called ``Hudson  effect,'' from the work of 
\citeauthor{hudson00}~\citeyear{hudson00}; e.g.\ \citeauthor{janse_et07}~\citeyear{janse_et07}, 
\citeauthor{shen_et12b}~\citeyear{shen_et12b}, \citeauthor{panesar_et13}~\citeyear{panesar_et13},
\citeauthor{wheatland_et18}~\citeyear{wheatland_et18}.)  To an
observer viewing from the upper left of the schematic, the measured intensity of the line-of-sight
component  of the foot of that positive-polarity field would increase between the times of Figures~7(a)
and 7(b).  Because the left side of the schematic  represents southeast in the solar images of this
paper, and because the region appears near the solar west limb viewed from HMI, the purple arrow
pointing  more toward the upper left in figure~7(b) compared to 7(a) is qualitatively consistent with
the  observed positive-flux increase during the time of minifilament eruption. This same process would
affect the negative-flux values also, but as mentioned earlier, we do not measure the negative flux
here because it is difficult to isolate. While further investigation  of this idea is beyond the scope
of the current paper, this scenario plausibly explains how there could be the observed increase in the
line-of-sight positive flux starting at the time of the jet onset shown in Figure~6.

Overall, we observe persistent flux cancelation over the time leading up to the jet.  Thus our 
findings are fully consistent with other recent studies showing that minifilaments that erupt to form
jets are frequently built by flux cancelation, and that continued cancelation triggers them to
erupt; see \citet{panesar_et16,panesar_et17,panesar_et18} for schematic illustrations on how this
build up and triggering might occur.  And see e.g., \citet{sterling_et15,sterling_et17} for schematic 
illustrations on how the erupting minifilament would form the jet.

EIS's field-of-view (Fig.~5(d); also see the white box in Fig.~1(i))  covers the location just to the east of the
interaction location of the erupting minifilament  and overlying field.  In Figure~5(d), the lineshifts labeled R2
are located near the AIA-flare-loop tops,  so if they were due to downflows from the tops of those loops, we would
expect a mixture of redshifts and  blueshifts; but we see only redshifts here.  Instead, the R2 redshifts  are
more consistent with material trapped in contracting reconnected closed field, producing ``supra-arcade
downflows,''  \citep[e.g.,][]{mckenzie00,savage_et11,warren_et11,savage_et12}, which have minimum velocities
near flare loop tops close to those observed here ($\sim$40~\kms).  Hence we expect the source of these redshifts
to be different from those seen  by EIS in AR loops \citep[e.g.,][]{delzanna08,doschek12}.

For the R1/B1-labeled red/blue pair of Doppler  shifts in Figure~5(d), which also
reach $\sim$40~\kms, Figure~3(e---f) and the accompanying video 
(and also other AIA videos from Fig.~1 showing the minifilament-eruption region close up) show that 
complex dynamics ensue when the minifilament pushes into the overlying horizontal field.  As mentioned in
the discussion of Figure~3 in \S\ref{subsec-images} above, the erupting minifilament shows
what appears to be untwisting motions (this would correspond to the minifilament flux rope with 
blue shading in Fig.~7(b)), along with northwestward-directed motions (green arrow 
in Figs.~3(e---f), and also southeastward-directed motions along the overlying horizontal field
(light-blue arrows in Figs.~3(d---f). Of course ``northwestward" and ``southeastward" only describe plane-of-sky
motions, but we could expect corresponding components either into or out of the plane of sky also.
EIS is apparently observing Doppler shifts from these opposite-directed motions; thus the shifts could be
from some combination between the possible untwisting motions \citep[as reported in, e.g.,][]{williams_et11} 
and the observed counterstreaming.  Due to the limited FOV coverage and the limited number of scans however, we are 
not able to specify 
more exactly what is the cause of the Doppler shifts (including for example, whether EIS is able
to resolve the red and blueshifts due to the untwisting motion alone).  Observations of similar events
with improved spectral coverage should clarify what occurs in situations such as these.


Recent models of the minifilament jet-producing mechanism 
(\citeauthor{wyper_et17}~\citeyear{wyper_et17}, \\
\citeauthor{wyper_et18}~\citeyear{wyper_et18}) argue that ``breakout
reconnection'' at a coronal magnetic null point between the pre-eruption field enveloping the 
minifilament and an ambient (nearly
vertical) coronal field is essential for initiating minifilament eruptions that produce single-spire jets. While this
may hold for the single-spire jets they modeled, initial breakout reconnection  appears not to be
essential in the two-sided-loop jet presented here; the minifilament flux-rope eruption undergoes clear
acceleration before the overlying field (where ``breakout'' reconnection eventually occurred at the interface) 
abruptly stops its
motion (Fig.~5).  Thus the possibility remains that tether-cutting \citep[e.g.][]{moore_et80}, or an
ideal MHD instability \citep[e.g.,][]{torok_et05}, prior to breakout might be the  primary
initiation mechanism for this minifilament eruption.

In summary, our view for this two-sided-loop jet is that (a) magnetic flux cancelation built
the minifilament field and triggered it to erupt, (b) the erupting minifilament field pushed up 
into largely horizontally oriented
pre-existing magnetic field, (c) the two-sided-loop jet spires developed due to reconnection 
between the erupting minifilament field and the overlying field.

Additionally, the central brightening in our event corresponds to
the JBP of single-spire jets, and is a miniature flare that accompanied the erupting
minifilament; it is analogous to large-scale flares accompanying 
typical-sized filament eruptions according to the standard flare model 
\citep[e.g.,][]{hirayama74,shibata_et95,moore_et01}.  Brightening of 
the horizontal jet loops would be due to a combination of heating of material at the reconnection 
location, density increase from evaporation from the near and far ends of the 
reconnected far-reaching loops, and density increase from leakage onto the horizontal segments
of material in the erupting minifilament field (including minifilament material
itself).

While there are now numerous examples of minifilament eruptions driving (single-spire) coronal jets 
\citep[e.g.,][]{shen_et12a,liu_et15,moore_et13,hong_et16,zhang_et16,sterling_et15,
panesar_et16}, it is necessary to study more than this single example of a two-sided-loop jet before we
might  conclude that two-sided-loop jets in general are also driven by minifilament eruptions, as other
recent  observation do not discuss erupting minifilaments in two-sided-loop jet formation
\citep{tian_et17,zheng_et18}.   Therefore more studies are needed to determine the most common cause of
two-sided-loop jets.

\acknowledgments

We thank an anonymous referee for comments that improved the paper.  This work was supported by funding 
from the Heliophysics Division of NASA's Science Mission Directorate
through the Heliophysics Guest Investigator (HGI)  Program, and the \hinode\ Project.  Data used in this
study were taken during \hinode\ 2015-2016 ``focused mode'' observations.  Hinode is a Japanese mission
developed and launched by ISAS/JAXA, with NAOJ as domestic partner and NASA and STFC (UK) as
international partners. It is operated by these agencies in co-operation with ESA and NSC (Norway).

\clearpage

\begin{figure}
\vspace{-1cm}
\hspace*{0.8cm}\includegraphics[angle=0,scale=0.80]{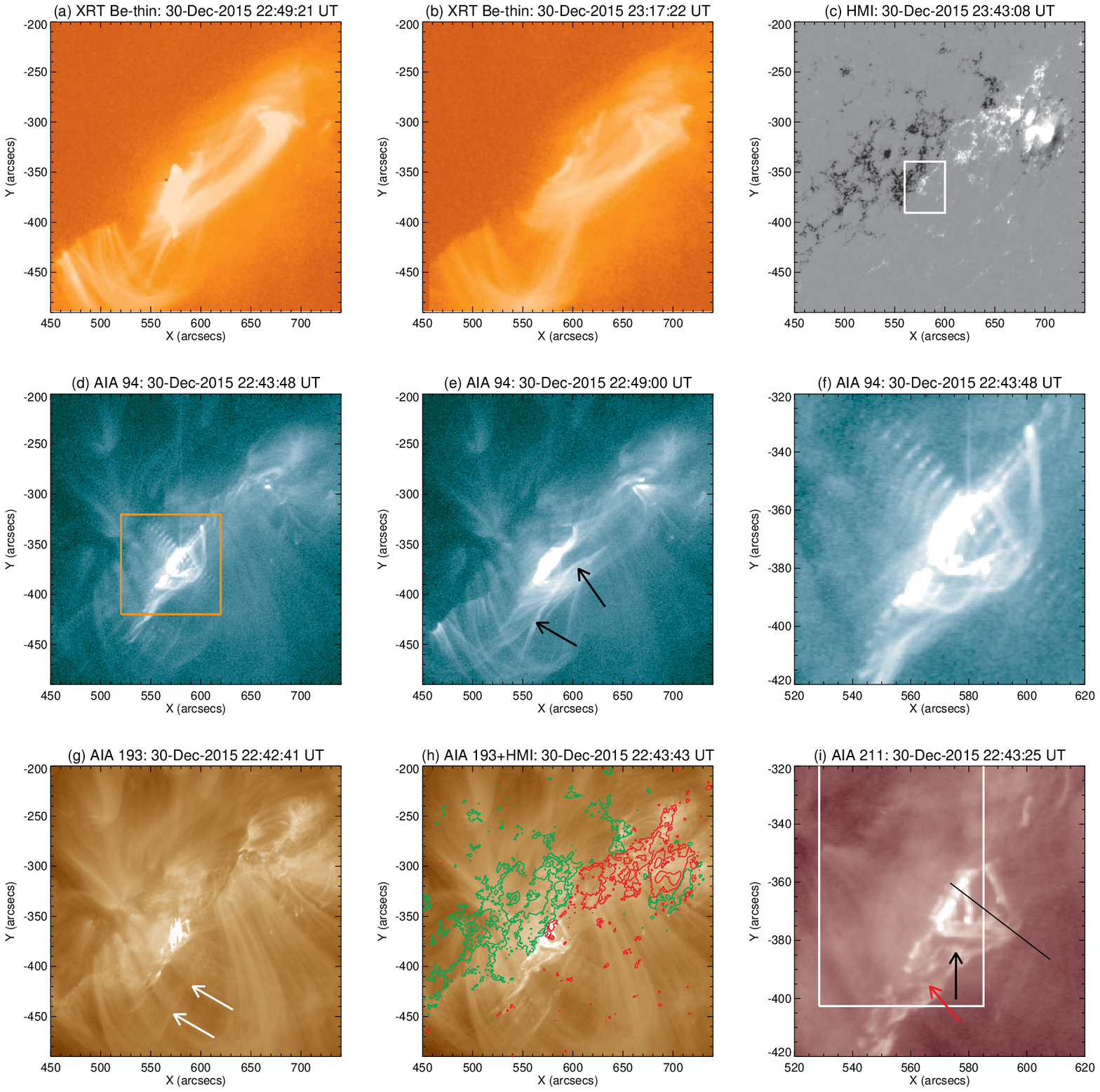}
\caption{\small Two-sided-loop jet images, and magnetic field. (a) and (b): From 
\hinode/XRT, showing the two-sided-loop jet from 
(a) near onset time, and (b) after peak intensity.  (c): HMI 
magnetogram of the jet region, with white (black) representing positive (negative) field.
The box shows the region over which 
magnetic flux changes were examined in Fig.~6. (d)---(f):  AIA 94~\AA\ images;
the box in (d) shows the region of intensity plotted in Fig.~4; arrows in (e) point to
oppositely directed flows along two jet-lobe strands. (g) and (h): AIA 193~\AA\ 
images. (i): An AIA~211~\AA\ image.  White arrows (g) show a magnetic loop segment 
situated nearly horizontally above the minifilament-eruption region.  In (i)
the erupting minifilament (black arrow) is just starting to strike that field, and produce the
two-sided-loop jet.  Contours of the magnetogram in (c) (at $\pm100, \pm750$~G) are overlaid 
in (h).  In (i) the white box shows the field of view of EIS (Fig.~5), and the black line is 
a fiducial showing the path over which the erupting minifilament's trajectory was measured for Fig.~4.
North is upward and west is to the right in this and all other solar figures in this paper.  
Videos corresponding to panels 1a\_1b, 1c, 1d\_1e, 1f, 1h, and 1i are available.}

\end{figure}
\clearpage

\begin{figure} 
\hspace*{-0.7cm}\includegraphics[angle=0,scale=1.0]{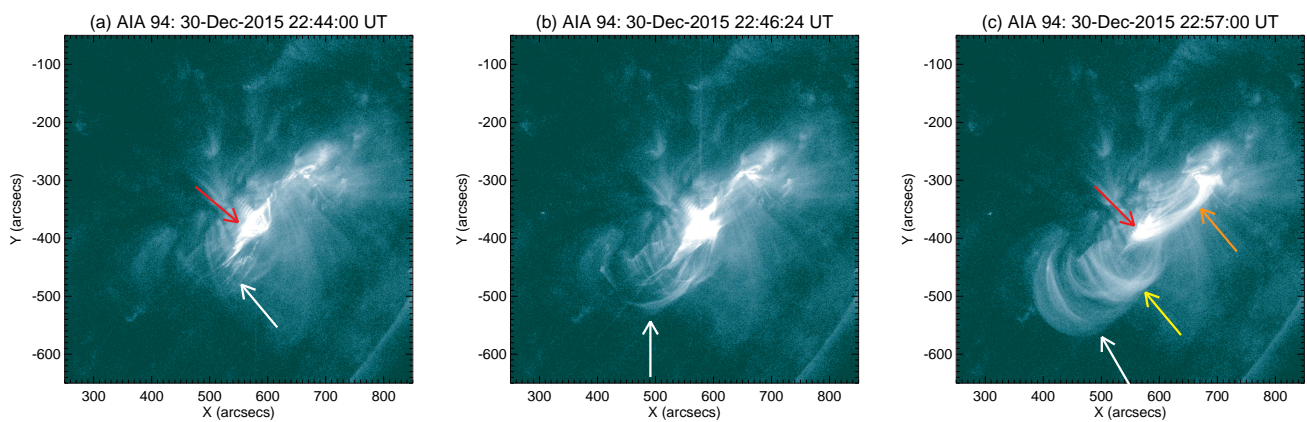}
\caption{\small AIA 94~\AA\ images showing larger-field-of-view perspective of the two-sided-loop jet of Fig.~1.  (a) After the initial 
brightening from the jet-initiation region (red arrow), a lobe of the jet starts to appear (white arrow).
(b) The lobe to the west continues to evolve toward the southeast (white arrow).  (c) The full structure of the two-sided-loop jet.
One lobe extends to the southeast, in this case along multiple field lines, with the white and yellow arrows pointing
to two such groups of field lines.  A second lobe extends to the northwest (orange arrow); this lobe 
was prominent from an earlier time in the jet's development in the X-ray images (Fig.~1a, 1b), but is just now becoming
apparent in 94~\AA\@.  Jets such as these are called two-sided-loop jets due to their roughly symmetric structure about 
the bright point at the initiation location (red arrow). Video 1d\_1e shows this sequence.}
\end{figure}
\clearpage

\begin{figure}
\epsscale{1.0}
\plotone{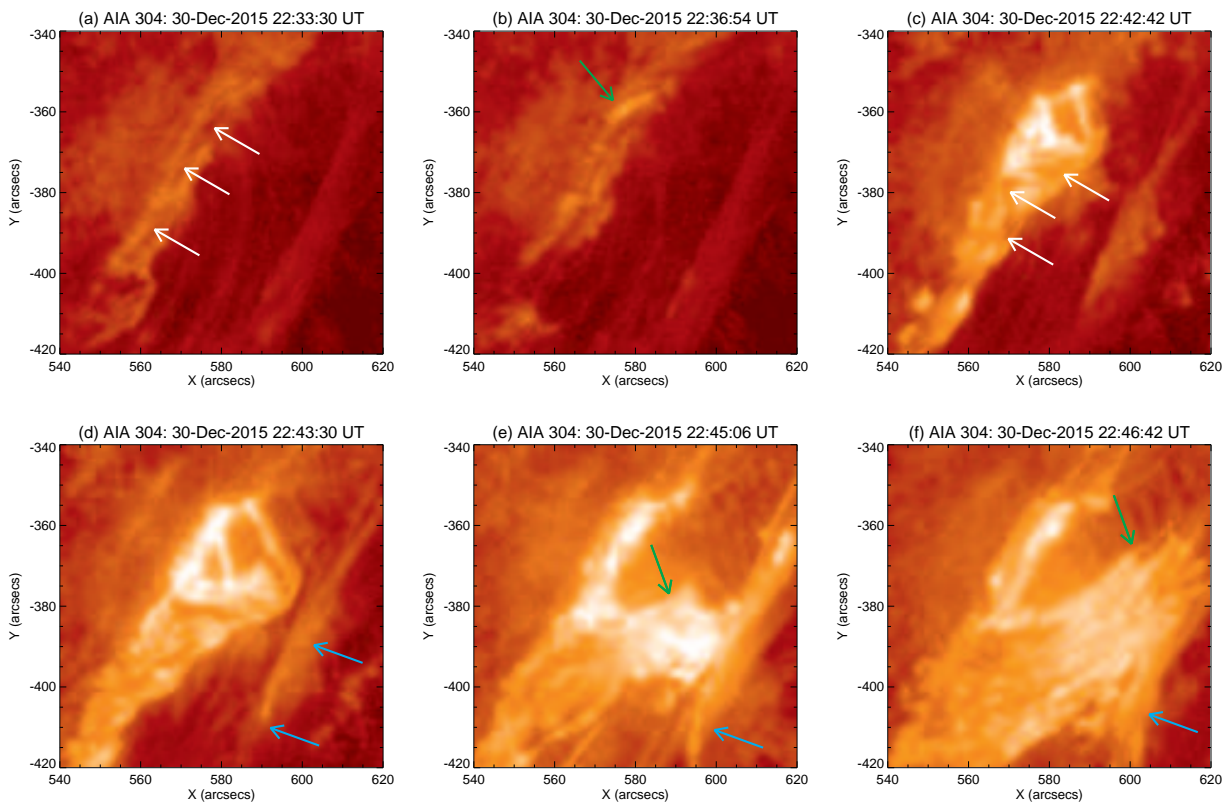}
\caption{\small Smaller FOV of the event of Fig~1, showing in AIA 304~\AA\ images the minifilament eruption 
leading to the two-sided-loop jet.  (a) Prior to the eruption the minifilament becomes visible very close to the surface 
(white arrows).  (b) A 
brightening (green arrow) becomes visible as the minifilament starts to lift off at the incipient stage of its 
eruption.  (c) As the eruption continues, the minifilament takes on a serpentine shape, indicative of writhing.
Brightenings continue to increase and also to spread along the length of the minifilament.  (d, e) As the
minifilament eruption continues, it impacts into overlying (presumably magnetic) structures that are oriented approximately
horizontal to the surface (light-blue arrows).  (f) Upon striking the overlying structure, the minifilament 
plasma is expelled in the northeast-southwest direction (green arrows in (e) and (f)); in addition, flows occur along 
the northwest-southeast-oriented field that is horizontal to the surface (light-blue arrows).  These motions are 
apparent in the accompanying video.}

\end{figure}
\clearpage

\begin{figure}
\epsscale{0.9}
\plotone{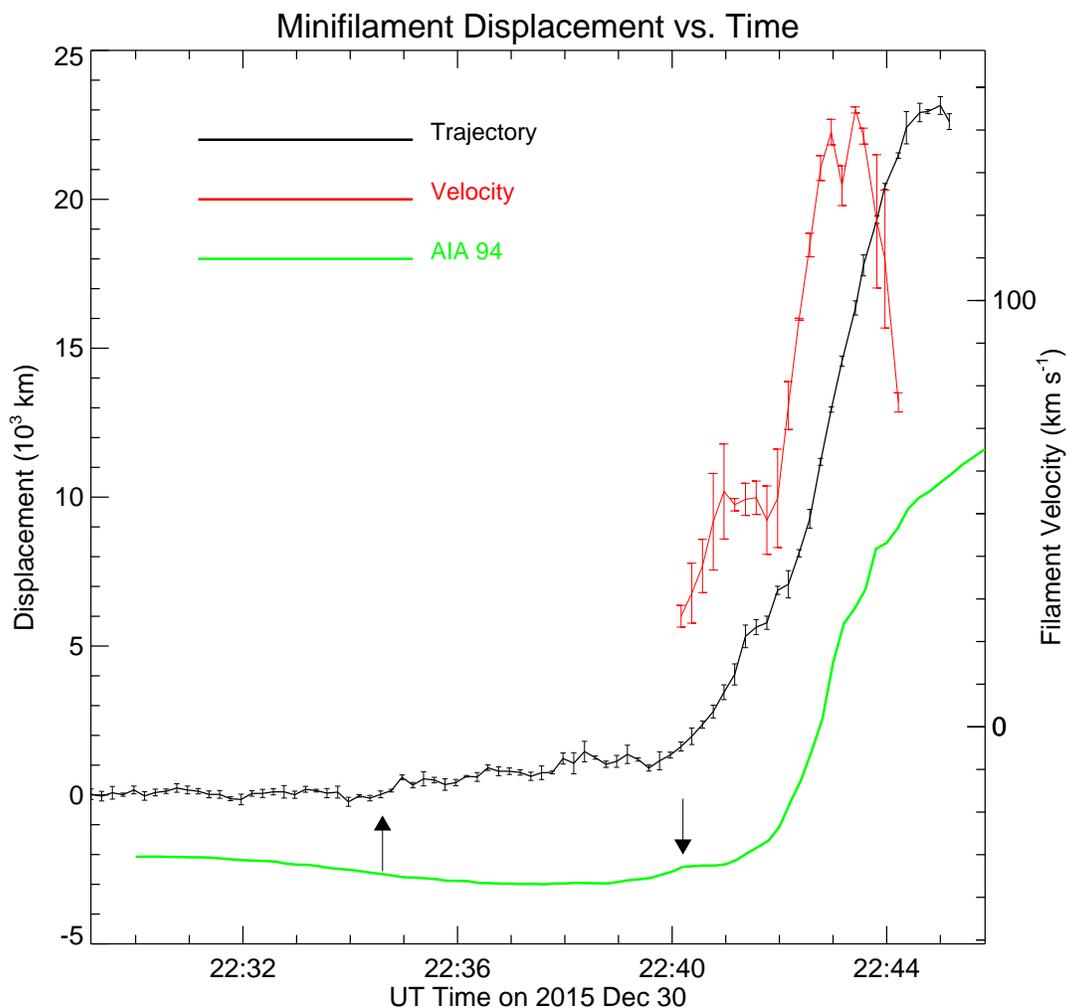}
\caption{\small Characteristics as functions of time of the minifilament that erupts and drives the two-sided-loop 
jet.  (Black line).  Height of top of minifilament segment along the path of the black fiducial line in Fig.~1(i). 
Error bars are 1$\sigma$ uncertainties
from three independent measurements.  (Orange) Velocities derived from the height trajectory (smoothed over four timesteps=48~s).  
(Green) Intensity from AIA 94~\AA\ channel, integrated over the region of the orange box in Fig.~1(d).  The arrow on the left 
shows the start of a pre-eruption rise in the minifilament's height.  The arrow on the right shows a minor peak in 94~\AA\ intensity, 
which coincides with onset 
of a faster rise; the 94~\AA\ video accompanying
Fig.~1 indicates a corresponding intensity brightening at the time for the left arrow also, but that peak in not visible in 
the green curve due to enhanced background emission.}
\end{figure}
\clearpage

\begin{figure} 
\hspace*{-0.7cm}\includegraphics[angle=0,scale=0.9]{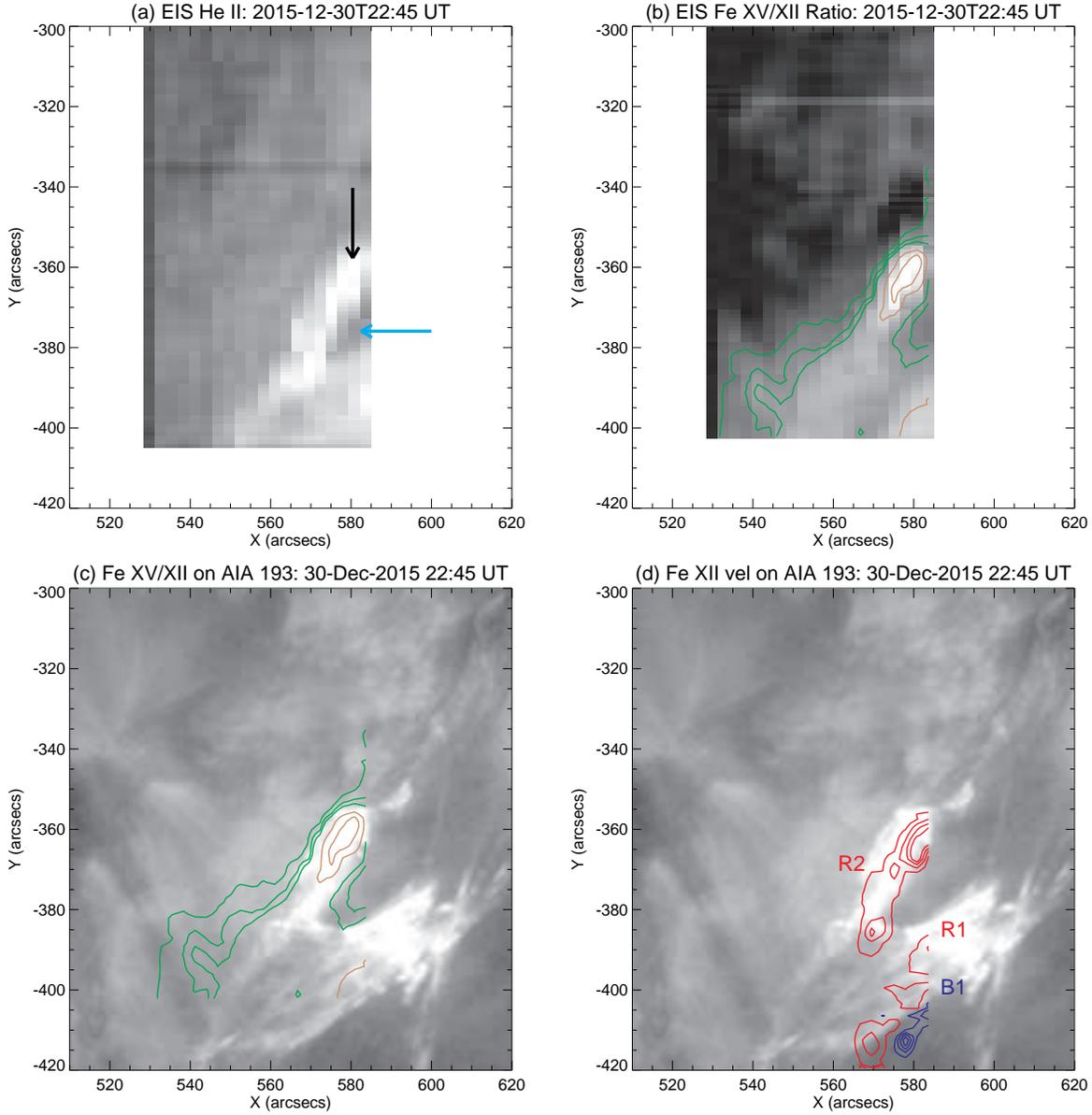}
\caption{\small \hinode/EIS observations.  (a) Raster image in \heii~304~\AA\ from the region of the white box in Fig.~1(i),
where the light-blue arrow shows a suspected void.
(b) Image of the ratio of \fexv-to-\fexii\ intensity rasters, providing an indication of relative temperatures, 
with bright (brown contours) and dark (green contours) locations respectively representing hotter and cooler 
regions.  (c) AIA 193~\AA\ image with counters of (b) overlaid. (d) AIA 193~\AA\ image with EIS \fexii\ Doppler 
contours at 10, 20, 30, and 40~\kms, with red/blue representing red/blue shifts. (Contours due to suspected noise 
have been removed from b--d.)}
\end{figure}
\clearpage

\begin{figure}
\hspace*{2.2cm}\includegraphics[angle=0,scale=0.60]{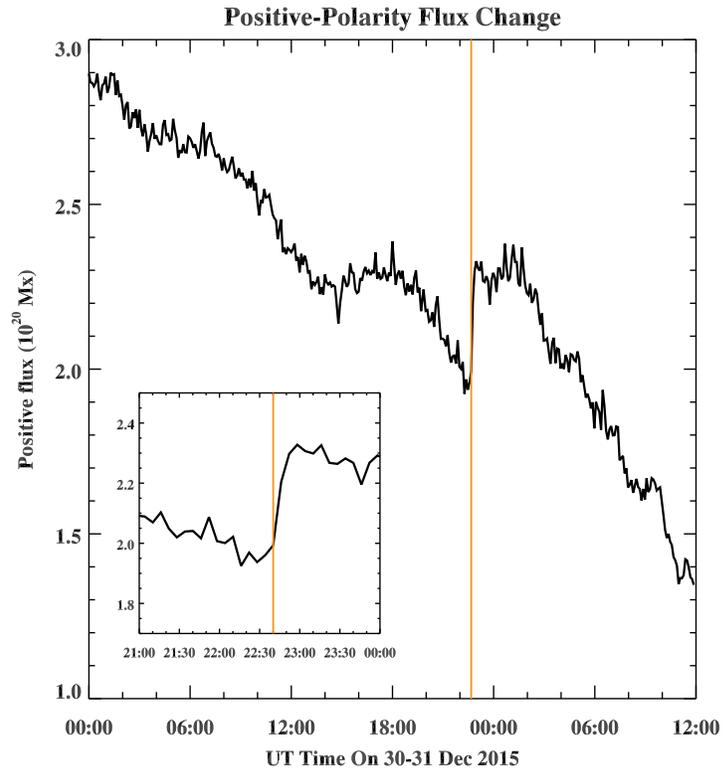}\vspace{1.0cm}
\caption{\small Positive magnetic flux, measured over the box of Fig.~1(c).
The orange line is at 22:40:08~UT, near the minifilament-eruption-onset time.  The 
insert plot is a closeup showing that the jump in flux value occurs very close to
the time of the jet.}

\end{figure}
\clearpage

\begin{figure}
\hspace*{2.2cm}\includegraphics[angle=270,scale=0.60]{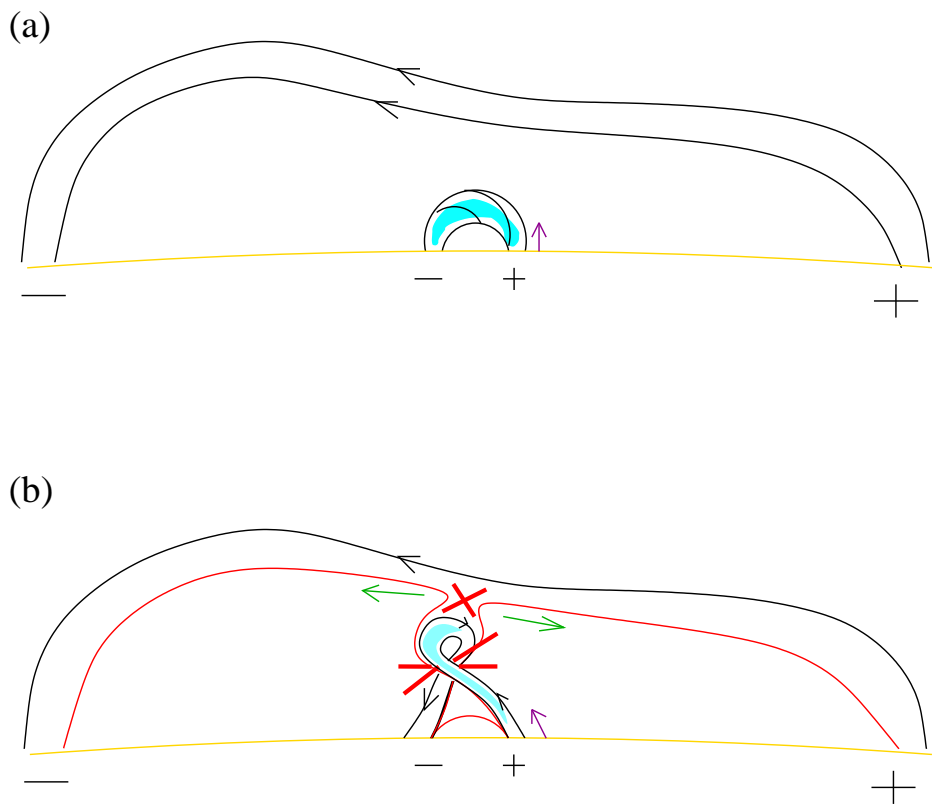}\vspace{1.0cm}
\caption{\small Schematic showing inferred production of the two-sided-loop jet.  (a) Setup prior to jet initiation, with the 
yellow curve representing the solar surface, black lines 
representing magnetic field lines, and the blue feature representing a cool minifilament suspended low inside of a twisted 
magnetic loop. Southeast is to the left and northwest to the right, in comparing with the observations in Figs.~1 and~5.  (b) Upon erupting, the 
minifilament field writhes enough to reconnect with the overlying field.  Red lines indicate reconnected field and
red X-es represent reconnection locations.  Reconnection also occurs between the legs of the 
writhing field, producing the red-semi-circular loop at the base; this is the brightening boxed in Fig.~1(d), and corresponds
to the JBP in the minifilament-eruption picture of \citet{sterling_et15}.  The green arrows represent flows in both directions
along the two-sided-loop jet.  The purple arrows show the orientation of the positive field direction at the surface and inside of the
erupting minifilament loop, with the arrow drawn outside of that loop for clarity.  If, for example, the eruption is observed
from the upper-left corner with a line-of-sight magnetograph, this positive component of the field will appear to be larger
in (b) than in (a), due to the field direction changing during the eruption; this change might explain the jump in measured positive magnetic flux at the time of the minifilament eruption displayed in Fig.~6.} 
\end{figure}


\begin{thebibliography}{}


\bibitem[Adams \etal(2014)]{adams_et14} Adams, M., Sterling, A. C., Moore, R. L., 
\& Gary, G. A.~2014, \apj, 783, 11

\bibitem[Alexander \& Fletcher(1999)]{alexander_et99} Alexander, D., \& Fletcher, L. 1999, \solphys, 
190, 167


\bibitem[Culhane \etal(2007)]{culhane_et07} Culhane, J. L., Harra, L. K., James, A. M., \etal~2007, 
\solphys, 243, 19

\bibitem[Cirtain \etal(2007)]{cirtain_et07} Cirtain, J. W., Golub, L., Winebarger, A. R., \etal~2007, Science, 318, 1580

\bibitem[Del~Zanna(2008)]{delzanna08} Del~Zanna, G. 2008, A\&A, 481, L49

\bibitem[Doschek \etal(2007)]{doschek_et07} Doschek, G. A., Mariska, J. T., Warren, H. P. \etal~2007, PASJ, 59, S707

\bibitem[Doschek(2012)]{doschek12} Doschek, G. A. 2012, \apj, 754, 153


\bibitem[Golub \etal(2007)]{golub_et07} Golub, L., Deluca, E., Austin, G., \etal~2007, \solphys, 243, 63


\bibitem[Green \etal(2018)]{green_et18} Green, L. M., T{\"o}r{\"o}k, T., Vr{\u s}nak, B., Manchester, W., \& Veronig, A.
2016, SSRv, 214, 46

\bibitem[Harra \etal(2017)]{harra_et17} Harra, L. K., Hara, H., Doschek, G. A., Matthews, S., Warren, H., Culhane, J. L., 
\& Woods, M. M. 2017, \apj, 842, 58

\bibitem[Hirayama(1974)]{hirayama74} Hirayama, T. 1974, \solphys, 34, 323

738L, 20

\bibitem[Hong \etal(2016)]{hong_et16} Hong, J., Jiang, Y., Yang, J., Yang, B., Xu, Z., \& Xiang, Y. 2016, \apj, 830, 60

\bibitem[Huang \etal(2012)]{huang_et12} Huang, Z., Madjarska, M. S., Doyle, J. G., \& Lamb, D. A. 2012, A\&A, 548,
A62

\bibitem[Hudson(2000)]{hudson00} Hudson, H. S. 2000, \apj, 531L, 75

\bibitem[Imada, Bamba, \& Kusano(2014)Imada~\etal]{imada_et14} Imada, S., Bamba, Y., \& Kusano, K. 2014, PASJ, 66, 17

\bibitem[Janse \& Low(2007)]{janse_et07} Janse, \AA. M., \& Low, B. C. 2007, A\&A, 472, 957

\bibitem[Jiang \etal(2013)]{jiang_et13} Jiang, Y., Bi, Y., Yang, J., Li, H., Yang, B., \& Zheng, R.~2013, \apj, 
775, 132

\bibitem[Lemen \etal(2012)]{lemen_et12} Lemen, J. R., Title, A. M., \& Akin, D. J.~\etal.~2012, 
\solphys, 275, 17

\bibitem[Li \etal(2015)]{li_et15} Li, X., Yang, S., Chen, H., \& Zhang, J. 2015 ApJ 814 13L

\bibitem[Liu \etal(2015)]{liu_et15} Liu, J., Wang, Y., Shen, C., Liu, K., Pan, Z., \& Wang, S. 2017, \apj, 813, 115

\bibitem[McCauley \etal(2015)]{mccauley_et15} McCauley, P. I., Su, Y. N., Schanche, N., Evans, K. E., Su, C.,
McKillop, S., \& Reeves, K. K. 2015, \solphys, 290, 1703



\bibitem[McKenzie(2000)]{mckenzie00} McKenzie, D. E. 2000, \solphys, 195, 381

\bibitem[Moore \& LaBonte(1980)]{moore_et80}Moore, R. L., and LaBonte, B.~1980, in Proc.\
Symp.\ on Solar and Interplanetary Dynamics, Reidel, Boston, 207

\bibitem[Moore \etal(1984)]{moore_et84} Moore, R. L., Hurford, G. J., Jones, H. P., \& Kane, S. R. 1984, \apj, 276, 379

\bibitem[Moore \etal(2001)]{moore_et01} Moore, R. L., Sterling, A. C., Hudson, H. S., \& Lemen, J. R. 2001, \apj, 552, 833


\bibitem[Moore \etal(2013)]{moore_et13} Moore R. L., Sterling A. C., Falconer D. A. \& Robe D. 2013  \apj, 769 134

\bibitem[Nistic\`o(2009)]{nistico_et09} Nistic\`o, G., Bothmer, V., Patsourakos, S., \& Zimbardo, G. 2009, Sol.\ Phys., 259, 87

\bibitem[Panesar \etal(2013)]{panesar_et13} Panesar, N. K., Innes, D. E., Tiwari, S. K, \& Low, B. C. 2013, 
A\&A, 549, 105

\bibitem[Panesar \etal(2016)]{panesar_et16} Panesar, N. K., Sterling, A. C., Moore, R. L., \& Chakrapani, P. 2016, \apjl, 832, L7



\bibitem[Panesar \etal(2017)]{panesar_et17}  Panesar, N. K., Sterling, A. C., \& Moore, \& R. L. 2017, \apj, 844, 131

\bibitem[Panesar \etal(2018)]{panesar_et18}  Panesar, N. K., Sterling, A. C., \& Moore, \& R. L. 2018, \apj, 853, 189


\bibitem[Parenti(2014)]{parenti14} Parenti, S. 2014, LRSP, 11, 1


\bibitem[Raouafi \etal(2016)]{raouafi_et16} Raouafi, N. E., Patsourakos, S., Pariat, E., \etal~2016, SSRv, 201, 1

\bibitem[Savage \& McKenzie (2011)]{savage_et11} Savage, S. L., \& McKenzie, D. E. 2011, \apj, 730, 98


\bibitem[Savage \etal(2012)Savage, McKenzie, \& Reeves(2012)]{savage_et12} Savage, S. L., McKenzie, D. E., \& Reeves, K. K. 
2012, \apjl, 747L, 40

\bibitem[Savcheva \etal(2007)]{savcheva_et07} Savcheva, A., Cirtain, J. W., DeLuca, E. E., \etal~2007, PASJ, 59S, 771S

\bibitem[Scherrer \etal(2012)]{scherrer_et12} Scherrer, P. H., \etal~2012, \solphys, 275,207

\bibitem[Shen \etal(2012a)]{shen_et12a}Shen, Y., Liu, Y. E., Su, J., \& Deng, Y. 2012a, \apj, 745, 164

\bibitem[Shen \etal(2012b)]{shen_et12b}Shen, Y., Liu, Y. E., Su, J., \& Deng, Y. 2012b, \apj, 750, 12

\bibitem[Shibata \etal(1992)]{shibata_et92}Shibata, K., Ishido, Y., Acton, L. W., \etal~1992, PASJ, 44L, 173

\bibitem[Shibata \etal(1994)]{shibata_et94}Shibata, K., Nitta, N., Matsumoto, R., Tajima, T., Yokoyama, T., Hirayama, T.,
\& Hudson, H.~1994, in 
Proc.\ of the International Symp.\ on the Yohkoh Scientific Results, X-Ray Solar Physics from
Yohkoh, ed.~Y. Uchida, T. Watanabe, K. Shibata, \& H. S. Hudson (Tokyo: Univ.\ Academy Press), 29

\bibitem[Shibata \etal(1995)]{shibata_et95}Shibata, K., Masuda, S., Shimojo, M., \etal~1995, \apj, 451L, 83

\bibitem[Shimojo \etal(1996)]{shimojo_et96}Shimojo, M., Hashimoto, S., Shibata, K., Hirayama, T., Hudson, H. S., 
\& Acton, L. W.~1996, PASJ, 48, 123


\bibitem[Sterling \& Moore(2005)]{sterling_et05} Sterling, A. C. \& Moore, R. L. 2005, \apj, 630, 1148


\bibitem[Sterling, Harra, \& Moore(2007)Sterling \etal]{sterling_et07} Sterling, A. C., Harra, L. K., 
\& Moore, R. L. 2007, \apj, 669, 1359




\bibitem[Sterling \etal(2014)]{sterling_et14} Sterling, A. C., Moore, R. L., Falconer, D. A., \& Knox, J. M. 2014, \apj, 788L, 20

\bibitem[Sterling \etal(2015)]{sterling_et15} Sterling, A. C., Moore, R. L., Falconer, D. A., \& Adams, M. 2015, Nature, 523, 437

\bibitem[Sterling \etal(2016)]{sterling_et16} Sterling, A. C., Moore, R. L., Falconer, D. A., \etal~2016, \apj, 821, 100

\bibitem[Sterling \etal(2017)]{sterling_et17} Sterling, A. C., Moore, R. L., Falconer, D. A., Panesar, N. K., \& Martinez, F. 2017, \apj, 844, 28

\bibitem[Sterling \etal(2018)Sterling, Moore, \& Panesar]{sterling_et18} Sterling, A. C., Moore, R. L., \& Panesar, N. K. 2018, 
\apj, in press

\bibitem[Tian \etal(2017)]{tian_et17} Tian, Z., Liu, Y., Shen, Y., Elmhamdi, A., Su, J., Liu, Y. D., \& Kordi, A. S.~2017,
ApJ, 845, 94.

\bibitem[T\"or\"ok \& Kliem(2005)]{torok_et05}T\"or\"ok, T.,  \& Kliem, B. 2005, \apjl, 630, L97

\bibitem[Wang \etal(1994)]{wang_et94} Wang, H., Ewell, M. W., Jr., Zirin, H., \& Ai, G. \apj, 424, 436

\bibitem[Warren \etal(2011)]{warren_et11} Warren, H. P., O'Brien, C. M., \& Sheeley, N. R. Jr. 2011, \apj, 742, 92

\bibitem[Wheatland \etal(2018)Wheatland, Melrose, \& Mastrano]{wheatland_et18} Wheatland, M. S., Melrose, D. B., \& Mastrano, A. 
2018, \apj, 864, 159

\bibitem[Williams \etal(2009)]{williams_et11} Williams, D. R., Harra, L. K., Brooks, D. H., Imada, S., \& Hansteen, 
V. H. 2011, \pasj, 61, 493

\bibitem[Wyper \etal(2017)]{wyper_et17} Wyper, P. F., Antiochos, S. K., \& DeVore, C. R. 2017, Nature, 544, 452


\bibitem[Wyper, DeVore, \& Antiochos(2018)Wyper \etal]{wyper_et18} Wyper, P. F., DeVore, C. R., \& Antiochos, S. K.~2018, \apj, 852, 98

\bibitem[Yokoyama \& Shibata(1995)]{yokoyama_et95}Yokoyama, T., \& Shibata, K. 1995, Nature, 375, 42


\bibitem[Yokoyama \& Shibata(1996)]{yokoyama_et96}Yokoyama, T., \& Shibata, K. 1996, PASJ, 48, 353


\bibitem[Young \& Muglach(2014)]{young_et14}Young, P. R., \& Muglach, K.~2014, \solphys, 289, 3313



\bibitem[Zhang \etal(2016)]{zhang_et16} Zhang, Q. M., Li, D., Ning, Z. J., Su, Y. N., Ji, H. S., \& Guy, Y. 2016,
\apj, 827, 27

\bibitem[Zheng \etal(2018)]{zheng_et18} Zheng, R., Chen, Y., Huang, Z., Wang, B., Song, H., \& Ning, H. 2018, \apj, 861, 108
\apj, 827, 27

\bibitem[Zwaan(1987)]{zwaan87}Zwaan, C. 1987, ARA\&A, 25, 83




\end{thebibliography}
\end{document}